\documentclass[pra,preprint,superscriptaddress,showpacs,preprintnumbers,eqsecnum,amsmath,amssymb]{revtex4}
\usepackage{graphicx}% Include figure files
\usepackage{dcolumn}% Align table columns on decimal point
\usepackage{bm}% bold math
\usepackage{epsfig}

\input xy %to make diagrams
\xyoption{all}

\begin{document}

\title{
Quantum Teleportation and Hidden Variables
}

\author{Amir Kalev}
\affiliation{Department of Physics, Technion - Israel Institute of
Technology, Haifa 32000, Israel}

\author{Sergio Rivera}
\affiliation{ Instituto de Ciencias Nucleares, Universidad
Nacional Aut\'{o}noma de M\'{e}xico, 01000 M\'{e}xico Distrito
Federal, M\'{e}xico }

\author{Pier A. Mello}
\affiliation{
Instituto de F\'{\i}sica, Universidad Nacional Aut\'{o}noma de
M\'{e}xico, 01000 M\'{e}xico Distrito Federal, M\'{e}xico
}

\date{
Jan. 18, 2008; Version: kalevArxiv}

%%%%%%%%%%%%%%%%%%%%%%%%%%%%%%%%%%%%%%%%%%
\begin{abstract}
In this paper we address the question as to what extent the
quantum-mechanical nature of the process is relevant for
teleportation of A spin-1/2 state. For this purpose we analyze the
possibility of underpinning teleportation with a
local-hidden-variable model. The nature of the models, which we
consider as legitimate candidates, guarantees the classical
character of all the probabilities which can be deduced from them.
When we try to describe the teleportation process following two
different mathematical routes, we find two different
hidden-variable densities, which thus end up having a doubtful
physical significance within the ``reality" that a hidden-variable
model tries to restore. This result we consider as a ``no-go
theorem" for the hidden-variable description of the teleportation
process. We also show that this kind of conflict arises when
considering successive measurements (one of which is selective
projective) for one spin-1/2 particle.
\end{abstract}
%%%%%%%%%%%%%%%%%%%%%%%%%%%%%%%%%%%%%%%%%%%%%%%%%%%%%%%%%%%%


\maketitle

\section{Introduction}
\label{intro}

The quantum teleportation (TP) process was designed to swap the
quantum-mechanical (QM) state  between two remote particles, to be
called 1 and 3, using the properties of entanglement
of particle 3 with an auxiliary particle 2 located in the vicinity of 1, plus
some information sent by classical means from the 1-2 pair to 3
\cite{bennet_et_al,vidman,braunstein_kimble}. In the so-called
``standard teleportation protocol" for spin-1/2 particles \cite{bennet_et_al}, particle 1 is initially in the state $|+\rangle^{\bm{n}}_1$
(defined to have spin projection $+$ in the direction of the unit vector $\bm{n}$),
while the 2-3 pair is in one of the maximally entangled states in the Bell basis.
A selective
projective measurement on the 1-2 pair of the appropriate observables defining a maximally entangled state in the Bell basis is performed.
The various outcomes of the 1-2 measurement are sent by some classical means to observer 3, who then selects the subensemble associated with one type of outcome.
The result is a projected state in which particle 3 is left in a state which coincides with $|+\rangle^{\bm{n}}_3$, or can be obtained from it by a simple unitary operation.

For a given quantum-mechanical process, the question as to what
extent the quantum nature of the phenomena involved is really
relevant is conceptually of fundamental importance. This question
has been  addressed in the literature for several, particular,
quantum processes. For example, it has been fully answered for
processes in systems of one spin-1/2 particle (see for example,
Refs. \cite{bell 1, mermin}): there are hidden-variable models
(HVMs) for one spin-1/2 particle that reproduce any expectation
value evaluated quantum mechanically. This is certainly true;
however, we shall study in Sec. \ref{succ_meas_one_s=1/2} below a
rather subtle situation that occurs when successive measurements
are concerned. For two spin-1/2 particles, the problem of
describing the quantum correlations with a HVM has been widely
studied (see, for instance, Refs. \cite{bell
1,mermin,fine,pitowsky}) and we shall not return to it here.
The present paper is concerned with the possibility of
underpinning the teleportation process -- which is a three
particles process -- with a HVM. This issue has been also analyzed
by several authors. For instance, we mention the study of
continuous-variables teleportation, in those cases that can be
expressed in terms of Wigner's functions \cite{braunstein_kimble}.
When all the Wigner functions involved in the protocol are
Gaussians, it was argued that they may be interpreted as providing
a local-hidden variable model (LHVM) -- where the phase-space
variables play the role of LHVs -- for the continuous-variables
teleportation protocol \cite{caves_wodkiewicz,kalev_et_al}.
As another example, Ref. \cite{zukowski} studies the
quantum-mechanical nature of
teleportation for spin-$1/2$ particles
 via the correlation
$E(\beta, \phi;\beta ', \phi')$
of certain observables defined for the 1-2 pair of particles and for particle 3.
Here, $\beta, \phi$ denote two angles
that determine the initial state of particle 1 to be teleported
(equivalent to specifying the unit vector ${\bm n}$ defined above),
and $\beta ', \phi'$ essentially determine the direction
in which the spin projection of particle 3 is measured.
The author studies whether that correlation can be modeled using a LHVM and finds a negative answer.

It is worth noticing that in Ref. \cite{kalev_et_al} the
subensemble described by a projection unto a definite maximally
entangled  state for the 1-2 pair is constructed, and its
consequences analyzed for the resulting marginal distribution for
particle 3, while in Ref. \cite{zukowski}  that projection is not
contemplated. It is precisely the rather deep effects of such a
projection for the full three-particle space that we wish to study
within a LHVM in the present paper.

In what follows we shall be able to make specific statements for a class of HVMs in which all the probabilities which can be computed are guaranteed to be of a classical nature, like those arising from ``balls in an urn" \cite{pitowsky}.
We shall find that, within these models, successive measurements
(the first one being
selective projective)
of commuting observables, such as those occurring in the TP problem, give rise to peculiar subtleties.

The paper is organized as follows. In the next section we
first review the standard TP process for
spin-1/2 particles; then, in Sec. \ref{hvm_TP}, we formulate a
LHVM for the process and exhibit the non trivial consequences of
having performed a selective measurement on the 1-2 pair and then
a measurement on particle 3. We shall find it instructive to
illustrate the consequences arising from successive measurements
in a much simpler problem: that of only one spin-$1/2$ particle;
this is done in Sec. \ref{succ_meas_one_s=1/2}. Finally, in Sec.
\ref{discussion_conclusions} we present a discussion of our
procedure and results.

\section{The standard teleportation protocol for spin-$1/2$ particles and a HV model}
\label{TPprocess}

\subsection{The quantum mechanical process}
\label{TP-QM}

Given two spin-1/2 particles, 1 and 2 say, we define the maximally entangled states in Bell's basis as
%%%%%%%%%%%%%%%%%%%%%%
\begin{subequations}
\begin{eqnarray}
|1, 1 \rangle_{12}^{z,x}
&=& \frac{1}{\sqrt{2}} \left[ | + \rangle_1^z | +  \rangle_2^z
+  | - \rangle_1^z | - \rangle _2^z \right]
\label{++}
\\
|1, -1 \rangle _{12}^{z,x}
&=& \frac{1}{\sqrt{2}}\left[ | +  \rangle_1^z | +  \rangle_2^z
- |-  \rangle_1^z | - \rangle _2 ^z\right]
\label{+-}
\\
|-1, 1 \rangle_{12}^{z,x}
&=& \frac{1}{\sqrt{2}}\left[ | +  \rangle_1^z | - \rangle_2 ^z
+  | - \rangle_1^z | +  \rangle _2^z \right]
\label{-+}
\\
|-1, -1 \rangle _{12}^{z,x}
&=& \frac{1}{\sqrt{2}}\left[ | +  \rangle_1^z | - \rangle_2 ^z
-  | - \rangle_1^z | +  \rangle _2^z \right] .
\label{__}
\end{eqnarray}
\label{bell_states}
\end{subequations}
%%%%%%%%%%%%%%%%%%%%%%
The notation $| + \rangle_1^z$ indicates the particle-1 state which is an eigenstate of $\sigma_{1z}$ with eigenvalue $+1$, and similarly for particle 2.
The Bell states of Eq. (\ref{bell_states}) are eigenstates of the two commuting operators
%%%%%%%%%%%%%%%%%%%%%%
\begin{subequations}
\begin{eqnarray}
\hat{B} &=&\sigma_{1z}\sigma_{2z}
\label{B}
\\
\hat{\bar{B}}&=&\sigma_{1x}\sigma_{2x},
\label{Bbar}
\end{eqnarray}
\label{B,Bbar}
\end{subequations}
%%%%%%%%%%%%%%%%%%%%%%
whose eigenvalues, that we shall call $\beta = \pm1$ and $\bar{\beta } = \pm1$, respectively, are indicated inside the kets on the LHS of Eq. (\ref{bell_states}).
The upper indices in these same kets remind us the axes that enter the definition of $B$ and $\bar{B}$.
In principle one could use an arbitrary pair of orthogonal directions to define these operators; however, for simplicity, the definite choice
specified in Eq. (\ref{B,Bbar}) will be adopted and will not be indicated any more in what follows.

In the standard teleportation protocol \cite{bennet_et_al} one starts out with three particles described by the state
%%%%%%%%%%%%%%%%%%%%%%
\begin{equation}
\left| \Psi \right\rangle_{123}
=  | + \rangle_1^{\bm{n}} \; |\beta_0  \bar{\beta}_0 \rangle_{23}.
\label{psi_123 1}
\end{equation}
%%%%%%%%%%%%%%%%%%%%%%
The state $| + \rangle_1^{\bm{n}}$ is the eigenstate of
$\bm{\sigma_1 \cdot n}$ with eigenvalue $+1$, for some
given
unit vector $\bm{n}$. The state for particles
$2,3$ in Eq. (\ref{psi_123 1}) is Bell's state
defined by the quantum numbers $\beta_0$, $\bar{\beta}_0$ [see Eq.
(\ref{bell_states})] for these particles. The state of Eq.
(\ref{psi_123 1}) can be expanded in terms of the complete set of
states (\ref{bell_states}) for particles 1 and 2, with the result
%%%%%%%%%%%%%%%%%%%%%%
\begin{equation}
\left| \Psi \right\rangle_{123}
= \sum_{\beta, \bar{\beta }}c_{\beta \bar{\beta }} \; |\beta \bar{\beta }\rangle_{12}
\; \left[ U_{\beta_0  \bar{\beta}_0}^{\beta \bar{\beta }}(3) \; | +\rangle_3^{\bm{n}}\right] ,
\label{psi_123 2}
\end{equation}
%%%%%%%%%%%%%%%%%%%%%%%%%
where $c_{\beta \bar{\beta }} = 1/2$ and
$U_{\beta_0  \bar{\beta}_0}^{\beta \bar{\beta }}(3)$ is a unitary transformation in the
space of particle 3 which depends on the pairs
$\beta_0 \bar{\beta }_0$, $\beta \bar{\beta }$ as follows
\cite{bennet_et_al}.
For $\{\beta_0  \bar{\beta}_0\}=\{1,1 \}$:
%%%%%%%%%%%%%%%%%%%%%%
\begin{subequations}
\begin{eqnarray}
U_{1,1}^{1,1}(3)
&=& I_3;
\hspace{0.5cm}
U_{1,1}^{1,-1}(3)
= \sigma_{3z};
\hspace{0.5cm}
U_{1,1}^{-1, 1}(3)
=\sigma_{3x};
\hspace{0.5cm}
U_{1,1}^{-1, -1}(3)
=
-i\sigma_{3y};
\label{U(beta_0=11)}
\end{eqnarray}
%%%%%%%%%%%%%%%%%%%%%%%%%%
For $\{\beta_0  \bar{\beta}_0\}=\{1,-1 \}$:
%%%%%%%%%%%%%%%%%%%%%%
\begin{eqnarray}
U_{1,-1}^{1,1}(3)
&=&
\sigma_{3z};
\hspace{0.5cm}
U_{1,-1}^{1,-1}(3)
= I_3;
\hspace{0.5cm}
U_{1,-1}^{-1, 1}
= i\sigma_{3y};
\hspace{0.5cm}
U_{1,-1}^{-1, -1}(3)
=-\sigma_{3x};
\label{U(beta_0=1,-1)}
\end{eqnarray}
%%%%%%%%%%%%%%%%%%%%%%%%%%
For $\{\beta_0  \bar{\beta}_0\}=\{-1,1 \}$:
%%%%%%%%%%%%%%%%%%%%%%
\begin{eqnarray}
U_{-1,1}^{1,1}(3)
&=& \sigma_{3x};
\hspace{0.5cm}
U_{-1,1}^{1,-1}(3)
= -i\sigma_{3y};
\hspace{0.5cm}
U_{-1,1}^{-1, 1}(3)
= I_3;
\hspace{0.5cm}
U_{-1,1}^{-1, -1}(3)
=\sigma_{3z};
\label{U(beta+0=-1,1)}
\end{eqnarray}
%%%%%%%%%%%%%%%%%%%%%%%%%%
For $\{\beta_0  \bar{\beta}_0\}=\{-1,-1 \}$:
%%%%%%%%%%%%%%%%%%%%%%
\begin{eqnarray}
U_{-1,-1}^{1,1}(3)
&=& -i \sigma_{3y};
\hspace{0.5cm}
U_{-1,-1}^{1,-1}(3)
= \sigma_{3x};
\hspace{0.5cm}
U_{-1,-1}^{-1, 1}(3)
= -\sigma_{3z};
\hspace{0.5cm}
U_{-1,-1}^{-1, -1}(3)
=-I_3  \; .
\nonumber \\
\label{U(beta_0=-1,-1)}
\end{eqnarray}
\label{U(beta)}
\end{subequations}
%%%%%%%%%%%%%%%%%%%%%%%%%%%%%%%%%%%%%%%%%%%%

Eq. (\ref{psi_123 2}) shows that in the three-particle state
$\left| \Psi \right\rangle_{123}$ the state
$|\beta \bar{\beta }\rangle_{12}$ is correlated with
the state of particle 3 which is obtained from a state identical to the original one of particle 1, i.e., $| + \rangle_3^{\bm{n}}$, transformed by the matrix
$U_{\beta_0  \bar{\beta}_0}^{\beta \bar{\beta }}(3)$.
The fundamental property of teleportation is that
the matrix $U_{\beta_0  \bar{\beta}_0}^{\beta \bar{\beta }}(3)$
is {\em independent of the vector} $\bm{n}$ that defines the original state
of particle 1, Eq. (\ref{psi_123 1}), to be teleported.
We remark that, up to a phase, the action of the unitary operator
$U_{\beta_0  \bar{\beta}_0}^{\beta \bar{\beta }}(3)$
on the state $|+ \rangle_3^{\bm{n}}$ has the influence of a rotation, to be called
$R_{\beta_0 \bar{\beta}_0}^{\beta \bar{\beta }}$, on the unit vector ${\bm{n}}$, i.e.
%%%%%%%%%%%%%%%%%%%%%
\begin{equation}
U_{\beta_0  \bar{\beta}_0}^{\beta \bar{\beta }}(3)
\; | + \rangle_3^{\bm{n}}
=e^{i\varphi _{\beta \bar{\beta }}} \;
| + \rangle_3^{R_{\beta_0 \bar{\beta }_0}^{\beta \bar{\beta }}\; \bm{n}} ,
\label{U_as_a_rotation}
\end{equation}
%%%%%%%%%%%%%%%%%%%%%
where $\varphi _{\beta \bar{\beta }}$ is a phase.
When the initial state is defined by $\{\beta_0  \bar{\beta}_0\}=\{-1,-1 \}$, i.e., for the singlet spin state, the rotation is given by the diagonal matrices
(only the diagonal matrix elements are indicated)
%%%%%%%%%%%%%%%%%%%%%%
\begin{equation}
R_{-1,-1}^{1,1} = (-1,1,-1);
\;
R_{-1,-1}^{1,-1} = (1,-1,-1);
\;
R_{-1,-1}^{-1,1} = (-1,-1,1);
\;
R_{-1,-1}^{-1,-1} = (1,1,1) \; ,
\label{R(beta)}
\end{equation}
%%%%%%%%%%%%%%%%%%%%%%%%%%
with similar results for the other three $\{\beta_0  \bar{\beta}_0\}$.

We now go back to Eq. (\ref{psi_123 2}).
In an ensemble of measurements of the observables $B, \bar{B}$ on particles 1,2 we obtain the result $\beta, \bar{\beta}$ with probability
$|c_{\beta \bar{\beta }}|^2 = 1/4$;
if 1-2 communicate --by classical means-- this result to observer 3, the latter can select the {\em sub-ensemble described by one term} in the state of
Eq. (\ref{psi_123 2})
(a selective projective measurement).
Observer 3 can then undo the transformation
$U_{\beta_0 \bar{\beta}_0}^{\beta \bar{\beta }}(3)$ and be left with the state $| + \rangle_3^{\bm{n}}$ for particle 3 which is identical to the state in which particle 1 originally was.
This is the essence of the so called ``standard teleportation protocol" \cite{bennet_et_al}.
Without such an information, observer 3 would have to use the full ensemble and, as is clear form Eq. (\ref{psi_123 1}) which factorizes the full state as that of particle 1 times that of particles 2 and 3, he would not be able to infer anything about the original state $|+ \rangle_1^{\bm{n}}$ of particle 1.

Suppose that observer 3 measures on particle 3 the expectation value of the observable $\hat{C} \equiv \bm{\sigma _3 \cdot c}$.
Here, the unit vector $\bm{c}$ gives the orientation of the Stern-Gerlach magnet used by
observer 3 in his measurement.
As explained above, no such measurement performed on the original state
of Eqs. (\ref{psi_123 1}) and (\ref{psi_123 2})
can give any information on the state $|+ \rangle_1^{\bm{n}}$ in which particle 1 was prepared, the situation being very different, however, if the state on which the measurement on 3 is performed is the {\em projection of one term} out of the original state (\ref{psi_123 2}).
The expectation value of the operator $\hat{C}$ in the projected state referred to above is
%%%%%%%%%%%%%%%%%%%%%
\begin{equation}
\langle\hat{C}\rangle_{QM}
=\frac{
_{123}\left\langle \Psi \right|
\hat{P}_{\beta \bar{\beta}}(12)
(\bm{\sigma _3 \cdot c})
\hat{P}_{\beta \bar{\beta}}(12)
\left| \Psi \right\rangle_{123}
}
{_{123}\left\langle \Psi \right|
\hat{P}_{\beta \bar{\beta}}(12)
\left|\Psi  \right\rangle_{123}
} \; ,
\label{av_in_proj_state 1}
\end{equation}
%%%%%%%%%%%%%%%%%%%%%
where
%%%%%%%%%%%%%%%%%%%%%
\begin{equation}
\hat{P}_{\beta \bar{\beta}}(12)
= |\beta \bar{\beta} \rangle_{12} \; _{12}\langle \beta \bar{\beta} |
\label{P_beta,betabar}
\end{equation}
%%%%%%%%%%%%%%%%%%%%%
is the projection operator unto the 1-2 state
$|\beta \bar{\beta} \rangle_{12}$.
We notice from Eq. (\ref{psi_123 2}) that the denominator appearing in
Eq. (\ref{av_in_proj_state 1}) is given by
%%%%%%%%%%%%%%%%%%%%%
\begin{equation}
_{123}\left\langle \Psi \right|
\hat{P}_{\beta \bar{\beta}}(12)
\left|\Psi  \right\rangle_{123}
\equiv Pr_{QM}(\beta \bar{\beta})
= |c_{\beta \bar{\beta}}|^2 =1/4.
\label{<P>}
\end{equation}
%%%%%%%%%%%%%%%%%%%%%
We shall compute the expectation value $\langle\hat{C}\rangle_{QM}$ of
Eq. (\ref{av_in_proj_state 1}) in a number of different ways and then find the HV version of each one of them.
The various mathematical routes to find the QM expectation value certainly give the same answer;
however, we shall see in Subsection \ref{hvm_TP} that requiring a similar equivalence for the corresponding HV images leads to a more subtle conclusion.

We shall call {\em Route A} the evaluation of $\langle\hat{C}\rangle_{QM}$ of Eq. (\ref{av_in_proj_state 1}) regarding it as the expectation value of the observable
$\hat{P}_{\beta \bar{\beta}}(12) (\bm{\sigma _3 \cdot c})\hat{P}_{\beta \bar{\beta}}(12)$
in the state $\left|\Psi  \right\rangle_{123}$, divided by
$_{123}\left\langle \Psi \right|
\hat{P}_{\beta \bar{\beta}}(12)
\left|\Psi  \right\rangle_{123}$.
A HVM for the observable and the state will then be proposed in Subsection \ref{hvm_TP}.

We shall denote by {\em Route B} the evaluation of $\langle\hat{C}\rangle_{QM}$ of Eq. (\ref{av_in_proj_state 1}) regarding it as the expectation value of the observable $\bm{\sigma _3 \cdot c}$ in the state obtained by letting the projector
$\hat{P}_{\beta \bar{\beta}}(12)$ act on the state
$\left|\Psi  \right\rangle_{123}$, i.e.,
%%%%%%%%%%%%%%%%%%%%%
\begin{subequations}
\begin{eqnarray}
|\Psi'\rangle
&=& \frac{\hat{P}_{\beta \bar{\beta}}(12)\left|\Psi  \right\rangle_{123}}
{\sqrt{Pr(\beta \bar{\beta})}}
\label{proj_post Ba}
\\
&=&
| \beta \bar{\beta} \rangle _{12}
U_{\beta_0 \bar{\beta}_0}^{\beta \bar{\beta}}(3)
| + \rangle_3^{\bm{n}}
\\
&=&e^{i\varphi _{\beta \bar{\beta }}} \;
| \beta \bar{\beta} \rangle _{12}
| + \rangle_3^{R_{\beta_0 \bar{\beta }_0}^{\beta \bar{\beta }}\; \bm{n}}
\\
&=&
e^{i\varphi _{\beta \bar{\beta }}} \;
| \beta \bar{\beta} \rangle _{12}
| + \rangle_3^{\bm{n}_{\beta_0 \bar{\beta}_0}^{\beta \bar{\beta }}} ,
\label{proj_post Bd}
\end{eqnarray}
\label{proj_post B}
\end{subequations}
%%%%%%%%%%%%%%%%%%%%%
where we have defined [see Eq. (\ref{U_as_a_rotation})]
%%%%%%%%%%%%%%%%%%%%%
\begin{equation}
\bm{n}_{\beta_0 \bar{\beta}_0}^{\beta \bar{\beta }}
= R_{\beta_0 \bar{\beta}_0}^{\beta \bar{\beta }} \; \bm{n}.
\label{Rn}
\end{equation}
%%%%%%%%%%%%%%%%%%%%%%%%
We then find
%%%%%%%%%%%%%%%%%%%%%
\begin{subequations}
\begin{eqnarray}
\langle\hat{C}\rangle_{QM}
&=& \; _{\;\;\;\;\;\;\;\;3}^{\bm{n}_{\beta_0 \bar{\beta}_0}^{\beta \bar{\beta }}}
\langle + |
\; _{12}\langle \beta \bar{\beta} |
\bm{\sigma_3 \cdot c}
 |\beta \bar{\beta} \rangle _{12}
| + \rangle_3^{\bm{n}_{\beta_0 \bar{\beta}_0}^{\beta \bar{\beta }}}
\label{av_in_proj_state B a}
\\
&=& _{\;\;\;\;\;\;\;\;3}^{\bm{n}_{\beta_0 \bar{\beta}_0}^{\beta \bar{\beta }}}\langle + |
\bm{\sigma_3 \cdot c}
| + \rangle_3^{\bm{n}_{\beta_0 \bar{\beta}_0}^{\beta \bar{\beta }}}
\label{av_in_proj_state B b}
\\
&=& _3^{\bm{n}}\langle + |
\bm{\sigma_3 \cdot c_{\beta_0 \bar{\beta}_0}^{\beta \bar{\beta }}}
| + \rangle_3^{\bm{n}} \; ,
\label{av_in_proj_state B c}
\end{eqnarray}
\label{av_in_proj_state B}
\end{subequations}
%%%%%%%%%%%%%%%%%%%%%
where we have defined
%%%%%%%%%%%%%%%%%%%%%
\begin{equation}
\bm{c}_{\beta_0 \bar{\beta}_0}^{\beta \bar{\beta }}
= [R_{\beta_0 \bar{\beta}_0}^{\beta \bar{\beta }}]^{-1} \; \bm{c}.
\label{R_1c}
\end{equation}
%%%%%%%%%%%%%%%%%%%%%%%%%%%
Eq. (\ref{av_in_proj_state B c}) states that the result of measuring
the expectation value $\langle\hat{C}\rangle_{QM}$ of Eq. (\ref{av_in_proj_state 1}) is identical to
that obtained if observer 3 had measured the average of
$\bm{\sigma _3 \cdot c_{\beta_0 \bar{\beta}_0}^{\beta \bar{\beta }}}$
on the single-particle state $| + \rangle_3^{\bm{n}}$,
which is the {\em same state in which particle 1 was originally prepared}.
We recall that $\bm{c}$ is the original orientation of the Stern Gerlach magnet used by observer 3, while $\bm{c}_{\beta_0 \bar{\beta}_0}^{\beta \bar{\beta }}$ is a new orientation of the magnet, which observer 3 knows how to fix using the information received from 1-2 by classical means.
This is an alternative way of describing the teleportation process.
A HVM for the observable $\bm{\sigma _3 \cdot c}$ and the state (\ref{proj_post B})
will then be proposed in the following subsection.

\subsection{A HVM for the standard teleportation protocol}
\label{hvm_TP}

Here we shall propose a HVM for each one of the routes indicated above.
In the various cases we make the following assumptions:

a) In a HV space $\Lambda$, with $\bm{\lambda} \in \Lambda$, we define the HV ``value" of the various observables to be measured, or, in the nomenclature of Ref. \cite{fine}, the ``response function" (giving the $\bm{\lambda}$-determined responses to the measurement) of the observable, which takes on the eigenvalues of the QM observable in various domains of the HV space $\Lambda$.
The response functions will depend, in principle, on the ``settings" defining the observable and, at least initially in our discussion, on the state $\psi$.

b) We assume a normalized (non-negative) probability density
%%%%%%%%%%%%%%%%%%%%%%%%
\begin{equation}
\rho_{\psi}(\bm{\lambda})
\label{rho(lambda) TP}
\end{equation}
%%%%%%%%%%%%%%%%%%%%%%%
defined on $\Lambda$ and dependent, in general, on the state of the system $\psi$.

As we proceed we shall need, on physical grounds, to limit these dependences.
It is then conceivable that some QM expectation values will not be reproduced by the LHVM;
however, independently of this point, we shall exhibit
certain subtle facts that emerge in constructing a HVM for successive measurements, which signal a conflict between the TP process and a HVM endowed with the above properties.

%%%%%%%%%%%%%%%%%%%%%%%%%%%%%%%%%%%%%%%%%%
\subsubsection{Route A}
\label{route_A}

The state is given by Eq. (\ref{psi_123 1}) and is defined by the parameters $\bm{n}$ and $\beta_0 \bar{\beta}_0$. We shall thus write the HV probability density as
%%%%%%%%%%%%%%%%%%%%%%%%%%
\begin{equation}
\rho_{\bm{n},\beta_0 \bar{\beta}_0}(\bm{\lambda}).
\label{rho}
\end{equation}
%%%%%%%%%%%%%%%%%%%%%%%%

The observable is
$\hat{P}_{\beta \bar{\beta}}(12)
(\bm{\sigma _3 \cdot c})\hat{P}_{\beta \bar{\beta}}(12)$.
We make the following assignment of response functions:
%%%%%%%%%%%%%%%%%%%%%%%%%%
\begin{subequations}
\begin{eqnarray}
\bm{\sigma _3 \cdot c} &\Longrightarrow&
C_{\bm{n},\beta_0 \bar{\beta}_0}(\bm{\lambda}; \bm{c})
= C_{\beta_0 \bar{\beta}_0}(\bm{\lambda}; \bm{c}) = \pm 1
\label{C}
\\
\hat{P}_{\beta \bar{\beta}}(12) &\Longrightarrow&
\Pi_{\bm{n},\beta_0 \bar{\beta}_0} (\bm{\lambda};\beta \bar{\beta})
= 0, 1    \; ,
\label{Pi}
\end{eqnarray}
\label{C,Pi}
\end{subequations}
%%%%%%%%%%%%%%%%%%%%%%%%
which are allowed to depend on the ``settings" and, for the time being, on the state.
For the observable $\bm{\sigma _3 \cdot c}$ the settings is $\bm{c}$: it
specifies the orientation of the corresponding Stern-Gerlach magnet.
The observable $\hat{P}_{\beta \bar{\beta}}(12)$ is specified by the
orientations $z$ and $x$ in Eqs. (\ref{bell_states}) and (\ref{B,Bbar}) (defining the Bell states) which we agreed to keep fixed, and by the pair of indices
$\{ \beta, \bar{\beta} \}$ specifying the particular Bell state on which we are projecting: these are then the settings for this observable.
On the basis of {\em locality}, as defined by Bell {\cite{bell 1}}, $C$ for particle 3 is allowed to depend on its own setting $\bm{c}$, but not on the setting $\beta \bar{\beta}$ of the observable
$\Pi$
for the pair of particles 1 and 2, which may be spatially separated from 3.
Similarly, $\Pi$ is allowed to depend on its own setting only, i.e., on
$\beta \bar{\beta}$, and not on $\bm{c}$.
With regards to the state dependence, since $C$ is an observable associated with particle 3,
on physical grounds it will not be
allowed to depend on $\bm{n}$, which determines the state in which particle 1 was prepared, independently of particles 2 and 3
[see Eq. (\ref{psi_123 1})].

Since the two QM operators in (\ref{C,Pi}) {\em commute},
$[\bm{\sigma _3 \cdot c}, \hat{P}_{\beta \bar{\beta}}(12)]=0$,
we assume, as in Ref. \cite{mermin}, that {\em the response function associated with their product is the product of the individual response functions} given in (\ref{C,Pi}).

The simplest expectation value to be modelled with this HVM is that of Eq. (\ref{<P>}), i.e., the denominator appearing in
Eq. (\ref{av_in_proj_state 1}):
%%%%%%%%%%%%%%%%%%%%%%%%%%
\begin{subequations}
\begin{eqnarray}
Pr_{QM}(\beta \bar{\beta})
=_{123}\left\langle \Psi \right|
\hat{P}_{\beta \bar{\beta}}(12)
\left|\Psi  \right\rangle_{123}
\Longrightarrow
Pr_{HV}(\beta \bar{\beta})
= \int
\Pi_{\bm{n},\beta_0 \bar{\beta}_0} (\bm{\lambda};\beta \bar{\beta})
\rho_{\bm{n},\beta_0 \bar{\beta}_0}(\bm{\lambda})
d \bm{\lambda} .
\nonumber \\
\label{HVM Pr(beta)}
\end{eqnarray}
%%%%%%%%%%%%%%%%%%%%%%%%
Next, we model the full expression (\ref{av_in_proj_state 1}) for the expectation value of $\hat{C}$, i.e.,
%%%%%%%%%%%%%%%%%%%%%%%%%
\begin{eqnarray}
\langle\hat{C}\rangle_{QM}
&=&\frac{
_{123}\left\langle \Psi \right|
\hat{P}_{\beta \bar{\beta}}(12)
(\bm{\sigma _3 \cdot c})
\hat{P}_{\beta \bar{\beta}}(12)
\left| \Psi \right\rangle_{123}
}
{_{123}\left\langle \Psi \right|
P_{\beta \bar{\beta}}(12)
\left|\Psi  \right\rangle_{123}
}
\nonumber \\
\nonumber \\
&&\hspace{1cm}\Longrightarrow
\langle C \rangle_{HV}^{(A)}
= \frac{1}{Pr_{HV}(\beta \bar{\beta})}
\int
C_{\beta_0 \bar{\beta}_0}(\bm{\lambda}; \bm{c})
\Pi_{\bm{n},\beta_0 \bar{\beta}_0} (\bm{\lambda};\beta \bar{\beta})
\rho_{\bm{n},\beta_0 \bar{\beta}_0}(\bm{\lambda})
d \bm{\lambda} \; .
\nonumber \\
\label{HVM <C>}
\end{eqnarray}
\label{HVM Pr(beta),<C>}
\end{subequations}
%%%%%%%%%%%%%%%%%%%%%%%%
We can rewrite the RHS of the correspondence (\ref{HVM <C>}) as
%%%%%%%%%%%%%%%%%%%%%%%%%%%%
\begin{subequations}
\begin{eqnarray}
\langle C \rangle_{HV}^{(A)}
= \int
C_{\beta_0 \bar{\beta}_0}(\bm{\lambda}; \bm{c})
\rho_{\bm{n},\beta_0 \bar{\beta}_0}(\bm{\lambda}|\beta \bar{\beta})
d \bm{\lambda} \; ,
\label{HVM <C> 1 a}
\end{eqnarray}
%%%%%%%%%%%%%%%%%%%%%%%%%%%%
where we have defined the {\em conditional probability density}
%%%%%%%%%%%%%%%%%%%%%%%
\begin{eqnarray}
\rho_{\bm{n},\beta_0 \bar{\beta}_0}(\bm{\lambda}|\beta \bar{\beta})
\equiv \frac
{\Pi_{\bm{n},\beta_0 \bar{\beta}_0} (\bm{\lambda};\beta \bar{\beta})
\rho_{\bm{n},\beta_0 \bar{\beta}_0}(\bm{\lambda})}
{Pr_{HV}(\beta \bar{\beta})}\; ,
\label{conditional}
\end{eqnarray}
\label{HVM <C> 1}
\end{subequations}
%%%%%%%%%%%%%%%%%%%%%%%%%%%%
which is the original $\rho_{\bm{n},\beta_0 \bar{\beta}_0}(\bm{\lambda})$
{\em conditioned on} $\bm{\lambda}$ {\em belonging to the domain}
$\Lambda_{\beta \bar{\beta}}$ where
$\Pi_{\bm{n},\beta_0 \bar{\beta}_0} (\bm{\lambda};\beta \bar{\beta})$
of Eq. (\ref{Pi}), for a specific $\beta \bar{\beta}$, takes on the value 1.

The probability density
$\rho_{\bm{n},\beta_0 \bar{\beta}_0}(\bm{\lambda})$
needs to reflect the fact that the QM state we are dealing with,
Eq. (\ref{psi_123 1}), is factorized into a state for particle 1 and a state for the pair of particles 2,3.
In this paper we shall achieve this requirement by splitting the HV as
$\bm{\lambda} \equiv \{\bm{\lambda_1}, \bm{\lambda_2}, \bm{\lambda_3} \}$, so that
%%%%%%%%%%%%%%%%%%%%%%%%%%%%
\begin{subequations}
\begin{eqnarray}
\rho_{\bm{n},\beta_0 \bar{\beta}_0}(\bm{\lambda})
=\rho_{\bm{n}}(\bm{\lambda_1})\rho_{\beta_0 \bar{\beta}_0}
(\bm{\lambda_2}, \bm{\lambda_3})\; .
\label{rho(123)}
\end{eqnarray}
%%%%%%%%%%%%%%%%%%%%%%%%%%%%
Similarly, we write for the response functions (\ref{C,Pi})
%%%%%%%%%%%%%%%%%%%%%%%%%%%
\begin{eqnarray}
C_{\beta_0 \bar{\beta}_0}(\bm{\lambda}; \bm{c})
&=& C_{\beta_0 \bar{\beta}_0}(\bm{\lambda_3}; \bm{c})
\label{C(3)}
\\
\Pi_{\bm{n},\beta_0 \bar{\beta}_0} (\bm{\lambda};\beta \bar{\beta})
&=& \Pi_{\bm{n},\beta_0 \bar{\beta}_0}
(\bm{\lambda_1}, \bm{\lambda_2};\beta \bar{\beta})\; .
\label{Pi(1,2)}
\end{eqnarray}
\label{C(3),Pi(1,2)}
\end{subequations}
%%%%%%%%%%%%%%%%%%%%%%%%%%%%
Then the RHS of the correspondence (\ref{HVM Pr(beta)}) becomes
%%%%%%%%%%%%%%%%%%%%%%%%%
\begin{equation}
Pr_{HV}(\beta \bar{\beta})
= \int \int \int
\Pi_{\bm{n},\beta_0 \bar{\beta}_0}
(\bm{\lambda_1}, \bm{\lambda_2};\beta \bar{\beta})
\rho_{\bm{n}}(\bm{\lambda_1})\rho_{\beta_0 \bar{\beta}_0}
(\bm{\lambda_2}, \bm{\lambda_3})
d \bm{\lambda_1} d \bm{\lambda_2} d \bm{\lambda_3}
\label{HVM Pr(beta) 12}
\end{equation}
%%%%%%%%%%%%%%%%%%%%%%
and the expectation value $\langle C \rangle_{HV}^{(A)}$ of
Eq. (\ref{HVM <C> 1 a}) becomes
%%%%%%%%%%%%%%%%%%%%%%%%%
\begin{equation}
\langle C \rangle_{HV}^{(A)}
= \int \int \int
C_{\beta_0 \bar{\beta}_0}(\bm{\lambda_3}; \bm{c})
\rho_{\bm{n},\beta_0 \bar{\beta}_0}(\bm{\lambda_1}, \bm{\lambda_2}, \bm{\lambda_3} |\beta \bar{\beta})
d \bm{\lambda_1} d \bm{\lambda_2} d \bm{\lambda_3} \;,
\label{HVM <C> 123}
\end{equation}
%%%%%%%%%%%%%%%%%%%%%%%%%%%%
where the conditional probability density is now
%%%%%%%%%%%%%%%%%%%%%%%
\begin{equation}
\rho_{\bm{n},\beta_0 \bar{\beta}_0}(\bm{\lambda_1}, \bm{\lambda_2}, \bm{\lambda_3}|\beta \bar{\beta})
\equiv \frac
{\Pi_{\bm{n},\beta_0 \bar{\beta}_0}
(\bm{\lambda_1}, \bm{\lambda_2};\beta \bar{\beta})
\rho_{\bm{n}}(\bm{\lambda_1})\rho_{\beta_0 \bar{\beta}_0}
(\bm{\lambda_2}, \bm{\lambda_3})}
{Pr_{HV}(\beta \bar{\beta})}\; .
\label{conditional_123}
\end{equation}
%%%%%%%%%%%%%%%%%%%%%%%%%%%%
This is the original density of Eq. (\ref{rho(123)})
conditioned on $\bm{\lambda_1}\bm{\lambda_2}$ belonging to the domain
$\Lambda_{\beta \bar{\beta}}$ where
$\Pi_{\bm{n},\beta_0 \bar{\beta}_0} (\bm{\lambda_1},\bm{\lambda_2};\beta \bar{\beta})$
of Eq. (\ref{Pi(1,2)}), for a specific $\beta \bar{\beta}$, takes on the value 1.
%%%%%%%%%%%%%%%%%%%%%%%%%%%%
\subsubsection{Route B}
\label{route_B}

The state is given by Eq. (\ref{proj_post B}) and is defined by the parameters $\beta \bar{\beta}$ and
$\bm{n}_{\beta_0 \bar{\beta}_0}^{\beta \bar{\beta}}$.
We shall thus write the HV probability density as
%%%%%%%%%%%%%%%%%%%%%%%%%%%%%%%
\begin{equation}
\rho_{\beta \bar{\beta}, \bm{n_{\beta_0 \bar{\beta}_0}^{\beta \bar{\beta }}}}(\bm{\lambda})
=\rho_{\beta \bar{\beta}}(\bm{\lambda_1}, \bm{\lambda_2})
\rho_{\bm{n_{\beta_0 \bar{\beta}_0}^{\beta \bar{\beta }}}}(\bm{\lambda_3})
\; .
\label{rho(B)}
\end{equation}
%%%%%%%%%%%%%%%%%%%%%%%%%%%%%%%
To construct the HV densities appearing on the RHS of (\ref{rho(B)})
we use the same set of rules that gave rise to the HV densities appearing on the RHS of (\ref{rho(123)}).

To the observable, which is now $(\bm{\sigma _3 \cdot c})$,
we assign the response function
%%%%%%%%%%%%%%%%%%%%%%%%%%%%%%%
\begin{equation}
\bm{\sigma _3 \cdot c} \Longrightarrow
C_{\beta \bar{\beta}, \bm{n_{\beta_0 \bar{\beta}_0}^{\beta \bar{\beta }}}}
(\bm{\lambda}; \bm{c})
%%%%%%%%%%%%%%%%
= C_{\bm{n_{\beta_0 \bar{\beta}_0}^{\beta \bar{\beta }}}}
(\bm{\lambda_3}; \bm{c})
%%%%%%%%%%%%%%%%%%%5
= \pm 1 \;
\label{c(B)}
\end{equation}
%%%%%%%%%%%%%%%%%%%%%%%%%%%%%%%
which, being an observable for particle 3, is allowed to depend
{\em only} on the parameters that define the state for particle 3 in the factorized state of Eq. (\ref{proj_post B}).
Notice that we are allowing
for a state dependence of the response function which, in principle, can make the function in Eq. (\ref{c(B)}) different from that of
Eq. (\ref{C(3)}). We come back to this point below.

We then model the expectation value of Eq. (\ref{av_in_proj_state B a})
as
%%%%%%%%%%%%%%%%%%%%%%%%%%%%%%%
\begin{eqnarray}
\langle C \rangle_{HV}^{(B)}
=\int \int \int C_{\bm{n_{\beta_0 \bar{\beta}_0}^{\beta \bar{\beta }}}}(\bm{\lambda_3}; \bm{c})
\rho_{\beta \bar{\beta}}(\bm{\lambda_1}, \bm{\lambda_2})
\rho_{\bm{n_{\beta_0 \bar{\beta}_0}^{\beta \bar{\beta }}}}(\bm{\lambda_3})
d \bm{\lambda_1} d \bm{\lambda_2} d \bm{\lambda_3} \; .
\label{<C> B a}
\end{eqnarray}
%%%%%%%%%%%%%%%%%%%%%%%%%%%%%%%

After proposing a HVM for Routes A and B,
we now examine the response functions and the densities entering the integrands in the expectation values of Eqs. (\ref{HVM <C> 123}) and (\ref{<C> B a}) above.
The motivation is that, if we should find two different response functions or densities
by pursuing two different mathematical routes (for the same experimental arrangement), then, as explained below, these quantities would end up having a doubtful physical significance within the ``reality" that a HVM tries to restore.

\underline{Response functions}. First, we recall that $C_{{\rm state}}(\bm{\lambda};\bm{c})$ is an observable quantity, since it gives one of the individual results when measuring the observable $\hat{C}$: at a given point in the HV $\bm{\Lambda}$ space, i.e., for a given $\bm{\lambda}$,
$C_{{\rm state}}(\bm{\lambda};\bm{c})$ takes on the value $1$ or the value $-1$,
and this is what shows up as an individual result of the measurement.
Now, in the mathematical representation of the problem we expect this value to be independent of the route that was followed to construct the HVM;
otherwise, the ``preassigned" value of the observable would be different in the two routes.
So far, $C_{\beta_0 \bar{\beta_0}}(\bm{\lambda_3};\bm{c})$ of
Eq. (\ref{C(3)})
was allowed to depend on the state $| \beta_0 \bar{\beta_0} \rangle_{23}$, and
$C_{\bm{n_{\beta_0 \bar{\beta}_0}^{\beta \bar{\beta }}}}(\bm{\lambda_3}; \bm{c})$ of Eq. (\ref{c(B)})
on the state
$|+\rangle^{\bm{n^{\beta \bar{\beta}}_{\beta_0 \bar{\beta}_0}}}_3$.
But then the two response functions are in general not equal: for instance, the second one depends on $\bm{n}$, while the first one does not.
For consistency, the two response functions
must be the same function of $\bm{\lambda_3}$
(and the setting $\bm{c}$) and this can only be achieved if they do not depend on the state, i.e., if we require
%%%%%%%%%%%%%%%%%%%%%%%%%%%%%%%
\begin{equation}
C_{\beta_0 \bar{\beta}_0}(\bm{\lambda_3}; \bm{c})
=C_{\bm{n_{\beta_0 \bar{\beta}_0}^{\beta \bar{\beta }}}}(\bm{\lambda_3}; \bm{c})
\equiv C(\bm{\lambda_3}; \bm{c}) \; .
\label{C=C'}
\end{equation}
%%%%%%%%%%%%%%%%%%%%%%%%5%%%%%%
As a result, {\em consistency of the notion of preassigned values requires the response functions not to depend on the state}.
This point will be illustrated in the simpler one-spin-$1/2$ case in the following section, in the discussion around Eq. (\ref{A indep of state}).
We may notice that the one-spin-$1/2$ model of App. \ref{model 1} does not fulfill this requirement, whereas the model of App. \ref{model 2} does.
We may add that though important the present conclusion is, it does not affect the second point to be discussed now.

\underline{HV densities}. Secondly, we compare the ``conditional" three-particle probability density
$\rho_{\bm{n},\beta_0 \bar{\beta}_0}(\bm{\lambda_1}, \bm{\lambda_2}, \bm{\lambda_3}|\beta \bar{\beta})$ of Eq. (\ref{conditional_123}) [occurring in the integrand for the expectation value (\ref{HVM <C> 123})], with the ``final" density
$\rho_{\beta \bar{\beta}, \bm{n_{\beta_0 \bar{\beta}_0}^{\beta \bar{\beta }}}}(\bm{\lambda})$ of
Eq. (\ref{rho(B)}) [entering the integrand for the expectation value
(\ref{<C> B a})].
The HV conditional density (\ref{conditional_123}) was obtained from a projection in  HV space unto the domain $\Lambda_{\beta \bar{\beta}}$.
In contrast, the final density (\ref{rho(B)}) was obtained by first projecting in Hilbert space the QM state vector unto $|\beta \bar{\beta} \rangle$ and then finding the HV image.

It is true that a HV density is not of the same nature as the HV response functions, in the sense that it is the latter which image the QM observables.
However, we should keep in mind that the very idea of a HV model is to restore the notion of reality, even though no effort is made to investigate the possible measurability of the density of such HV's.
Now, the two densities we mentioned in the previous paragraph were obtained simply by producing HVMs
of two mathematical routes that give the same QM result.
Should these densities turn out to be different, their physical significance, within the ``reality" that a HV model tries to restore, would be doubtful.

We now proceed to show that the two densities that we mentioned
are indeed
not the same. We first make a general remark. The QM
Bell states of Eq. (\ref{bell_states}) are either symmetric or
antisymmetric with respect to an interchange of the two particles,
so that the resulting density matrix is symmetric with respect to
the same operation. We expect the HV density associated with such
QM states to reflect this fact: $\rho_{\beta_0
\bar{\beta}_0}(\bm{\lambda}_2, \bm{\lambda}_3) = \rho_{\beta_0
\bar{\beta}_0}(\bm{\lambda}_3, \bm{\lambda}_2)$ in Eq.
(\ref{rho(123)}) and $\rho_{\beta \bar{\beta}}(\bm{\lambda}_1,
\bm{\lambda}_2) = \rho_{\beta \bar{\beta}}(\bm{\lambda}_2,
\bm{\lambda}_1)$ in Eq. (\ref{rho(B)}).
Notice also that
$\hat{P}_{\beta_0 \bar{\beta}_0}(3, 2)
=\hat{P}_{\beta_0 \bar{\beta}_0}(2,3)$
and that the HV image should reflect this symmetry.

Consider now the case $\beta=\beta_0$,
$\bar{\beta}=\bar{\beta}_0$. If the conditional and final
densities were the same, we would have
%%%%%%%%%%%%%%%%%%%%%%%%%%%%%%%
\begin{equation}
\Pi(\bm{\lambda_1}, \bm{\lambda_2};\beta_0 \bar{\beta}_0)
\rho_{\bm{n}}(\bm{\lambda_1}) \rho_{\beta_0
\bar{\beta}_0}(\bm{\lambda_2}, \bm{\lambda_3})=p\,\rho_{\beta_0
\bar{\beta}_0}(\bm{\lambda_1}, \bm{\lambda_2})
\rho_{\bm{n}}(\bm{\lambda_3})\;, \;\;\;\;\;\;
\forall \bm{\lambda_1},  \bm{\lambda_2}, \bm{\lambda_3} \; ,
\label{C=F}
\end{equation}
%%%%%%%%%%%%%%%%%%%%%%%%%%%%%%
where
%%%%%%%%%%%%%%%%%%%%%%%%%%%%%%%
\begin{equation}\label{pHV}
p=
\int \Pi(\bm{\lambda_1'}, \bm{\lambda_2'};\beta_0
\bar{\beta}_0) \rho_{\bm{n}}(\bm{\lambda_1'}) \rho_{\beta_0
\bar{\beta}_0}(\bm{\lambda_2'}, \bm{\lambda_3'})
d\bm{\lambda_1'}d\bm{\lambda_2'}d\bm{\lambda_3'}
=\frac{1}{4} \; .
%\label{p 0}
\end{equation}
%%%%%%%%%%%%%%%%%%%%%%%%%%%%%%
Here, the single-particle HV density on the LHS of Eq. (\ref{C=F}) is the same
function as that on the RHS of the equation, when $\bm{\lambda_1}$
is replaced by $\bm{\lambda_3}$, because they are the HV image of
the same physical QM state. Similarly, and for the same reason,
the two-particle HV densities on the two sides of the equation are
the same function when changing $\bm{\lambda_2}, \bm{\lambda_3}$
on the LHS to $\bm{\lambda_1}, \bm{\lambda_2}$ on the RHS. Also,
we have used the fact that $\bm{n_{\beta_0 \bar{\beta}_0}^{\beta_0
\bar{\beta }_0}}=\bm{n}$, as it can be seen from Eqs.
(\ref{U(beta)})-(\ref{R(beta)}) and (\ref{Rn}) for
$\beta=\beta_0$, $\bar{\beta}=\bar{\beta}_0$.

If Eq. (\ref{C=F}) is valid $\forall \bm{\lambda_1}, \bm{\lambda_2}, \bm{\lambda_3}$,
it should also hold if we interchange the values of the two variables
$\bm{\lambda_1}$ and $\bm{\lambda_3}$, i.e.
$\bm{\lambda_1} \leftrightarrow \bm{\lambda_3}$.
Noticing from  Eq. (\ref{pHV}) that $p=1/4$ does not depend on $\bm{\lambda_1}$ nor on
$\bm{\lambda_3}$, and using the symmetry properties discussed above Eq. (\ref{C=F}),
we obtain
%%%%%%%%%%%%%%%%%%%%%%%%%%%%%%
\begin{equation}
\Pi(\bm{\lambda_2}, \bm{\lambda_3};\beta_0 \bar{\beta}_0)
\rho_{\bm{n}}(\bm{\lambda_3})\rho_{\beta_0
\bar{\beta}_0}(\bm{\lambda_1}, \bm{\lambda_2})=p\,\rho_{\beta_0
\bar{\beta}_0}(\bm{\lambda_2},
\bm{\lambda_3})\rho_{\bm{n}}(\bm{\lambda_1})\,. \label{C=F_1}
\end{equation}
%%%%%%%%%%%%%%%%%%%%%%%%%%%%%
For $p\neq 0$ we solve  Eq. (\ref{C=F}) for $\rho_{\beta_0
\bar{\beta}_0}(\bm{\lambda_1}, \bm{\lambda_2})
\rho_{\bm{n}}(\bm{\lambda_3})$ and substitute it  in
(\ref{C=F_1}); we obtain
%%%%%%%%%%%%%%%%%%%%%%%%%%%%%
\begin{equation}
\Pi(\bm{\lambda_2}, \bm{\lambda_3};\beta_0 \bar{\beta}_0)
\Pi(\bm{\lambda_1}, \bm{\lambda_2};\beta_0 \bar{\beta}_0)
\rho_{\bm{n}}(\bm{\lambda_1}) \rho_{\beta_0
\bar{\beta}_0}(\bm{\lambda_2},
\bm{\lambda_3})=p^2\,\rho_{\bm{n}}(\bm{\lambda_1})\rho_{\beta_0
\bar{\beta}_0}(\bm{\lambda_2}, \bm{\lambda_3}) \,. \label{C=F_2}
\end{equation}
%%%%%%%%%%%%%%%%%%%%%%%
But $\Pi(\bm{\lambda_2}, \bm{\lambda_3};\beta_0
\bar{\beta}_0)\rho_{\beta_0 \bar{\beta}_0}(\bm{\lambda_2},
\bm{\lambda_3})\rho_{\bm{n}}(\bm{\lambda_1})
=\rho_{\beta_0 \bar{\beta}_0}(\bm{\lambda_2},
\bm{\lambda_3})\rho_{\bm{n}}(\bm{\lambda_1})$, as it can be seen multiplying both sides of
(\ref{C=F_1}) by $\Pi(\bm{\lambda_2}, \bm{\lambda_3};\beta_0
\bar{\beta}_0)$, using the fact that $\Pi(\bm{\lambda_2},
\bm{\lambda_3};\beta_0 \bar{\beta}_0)$ is idempotent
and that $p \neq 0$.
So, (\ref{C=F_2}) becomes
%%%%%%%%%%%%%%%%%%%%%%%%%%%%%
\begin{equation}
\Pi(\bm{\lambda_1}, \bm{\lambda_2};\beta_0 \bar{\beta}_0)
\rho_{\bm{n}}(\bm{\lambda_1}) \rho_{\beta_0
\bar{\beta}_0}(\bm{\lambda_2},
\bm{\lambda_3})=p^2\,\rho_{\bm{n}}(\bm{\lambda_1})\rho_{\beta_0
\bar{\beta}_0}(\bm{\lambda_2}, \bm{\lambda_3}) \,. \label{C=F_3}
\end{equation}
%%%%%%%%%%%%%%%%%%%%%%%
Integrating both sides over $\bm{\lambda_1}$, $\bm{\lambda_2}$,
and $\bm{\lambda_3}$ we obtain
%%%%%%%%%%%%%%%%%%%%%%%%%%%%%
\begin{equation}
\int\Pi(\bm{\lambda_1}, \bm{\lambda_2};\beta_0 \bar{\beta}_0)
\rho_{\bm{n}}(\bm{\lambda_1}) \rho_{\beta_0
\bar{\beta}_0}(\bm{\lambda_2},
\bm{\lambda_3})d\bm{\lambda_1}d\bm{\lambda_2}d\bm{\lambda_3}=p^2
\,. \label{C=F_4}
\end{equation}
%%%%%%%%%%%%%%%%%%%%%%%
From (\ref{pHV}), the LHS is $p$, so that
%%%%%%%%%%%%%%%%%%%%%%%%%%%%%
\begin{equation}
p=p^2, \label{p=p^2TP}
\end{equation}
%%%%%%%%%%%%%%%%%%%%%%%
which holds only for $p=0$ or $p=1$; for $p =1/4$ we have a
contradiction.

We thus conclude that the Hilbert space projection operation cannot be described in terms of, and is in conflict with, the operation of projection in HV space.
I.e., application of the rules of classical statistics in HV space to the HV density in order to obtain a conditional probability, as in Route A, does not give the same result as projecting in Hilbert space first and then finding the HV model.
I.e., the two operations --projection and HV modeling-- cannot be interchanged.

In other words, the HV density for the state projected in Hilbert space cannot be obtained manipulating the HV density for the original state following the rules of statistics in HV space:
{\em the Hilbert space projection operation cannot be described in terms of, and is in conflict with, the standard rules of statistics in HV space}.

The main conclusion stated above is based on a number of assumptions that have been made in constructing our HVM, which we now list:

1) A {\em HV density} of the form (\ref{rho(lambda) TP}).

2) {\em Locality}, as defined by Bell \cite{bell 1}. This property was used in
Eq. (\ref{C,Pi}) to eliminate from the response function of one subsystem the dependence on the setting of the instrument used to make a measurement on another subsystem which may be spatially separated.

3) {\em Splitting of the HV} $\bm{\lambda}$ in three sets,
$\{\bm{\lambda}_1, \bm{\lambda}_2, \bm{\lambda}_3 \}$, to be associated with each of the three particles, respectively.
This feature was used starting from Eq. (\ref{rho(123)}) to represent states which, quantum-mechanically, are separable with respect to two subsystems.
This last
assumption  was needed
within the analysis we have presented here; it would be desirable to eliminate it, although
at this moment this is still an open question for the present authors.

We recall that a
consequence that emerges from our analysis is the
{\em independence of the response functions on the state} when using projectors in the formalism.
This property is needed for consistency of the HVMs obtained when the projector is considered as an observable or as acting on the state to produce a new state
[see Eq. (\ref{C=C'})].

%%%%%%%%%%%%%%%%%%%%%%%%%%%%%%%%%%
\section{Illustration of the conflict  with successive measurements for one spin-$1/2$ particle}
\label{succ_meas_one_s=1/2}

In this section we illustrate, in a simple one spin-$1/2$ problem, the conflict that was presented in the previous section for the more complex three-particle TP problem.

\subsection{The quantum mechanical problem of successive measurements for one spin-$1/2$
particle}

Consider the  observable defined in spin space as
%%%%%%%%%%%%%%%%%%%%%%%%
\begin{subequations}
\begin{eqnarray}
\hat{A}&=& \alpha_1 \bm{\sigma \cdot a} + \alpha_2
\label{A a}
\\
&=& \sum_{s=\pm} |s \rangle^{\bm{a}} \; a_s \; ^{\bm{a}}\langle s | ;
\hspace{1cm}{\rm Eigenvalues:}\hspace{.5cm}
a_{\pm}= \pm \alpha_1 +  \alpha_2\;.
\label{A b}
\end{eqnarray}
\label{A}
\end{subequations}
%%%%%%%%%%%%%%%%%%%%%%%%
Here, $\bm{\sigma }$ is the vector of Pauli matrices. The unit vector $\bm{a}$, which might represent the orientation of a Stern-Gerlach magnet, will be called the {\em instrument setting};
$\alpha_1$ and $\alpha_2$ are numerical constants. The symbol $s$ takes on the values $+$
(or $-$),  and the ket $|+(-) \rangle^{\bm{a}}$ indicates the state with spin up (down) in the direction
$\bm{a}$.
The observable is written in Eq. (\ref{A b}) in its spectral resolution which also indicates its eigenvalues.

The most general projector in spin space which commutes with the operator
$\hat{A}$ is
%%%%%%%%%%%%%%%%%%%%%%%%
\begin{equation}
\hat{P}^{\bm{a}}_{s_0}
=| s_0 \rangle^{\bm{a}} \; ^{\bm{a}}\langle s_0  |\;,
\label{B=P}
\end{equation}
%%%%%%%%%%%%%%%%%%%%%%%%
with $s_0 = \pm$; it projects unto the state $| s_0 \rangle^{\bm{a}}$.
We easily find
%%%%%%%%%%%%%%%%%%%%%%%%
\begin{equation}
\hat{P}_{\pm}^{\bm{a}} = \frac12 (\pm \bm{\sigma \cdot a} + 1).
\label{P+-}
\end{equation}
%%%%%%%%%%%%%%%%%%%%%%%%
The projector $\hat{P}^{\bm{a}}_{s_0}$ is chosen to commute with the observable $\hat{A}$
in order to have a closer analogy with the situation studied in the previous section.
However, notice that in Eq. (\ref{av_in_proj_state 1}) the projector
$P_{\beta \bar{\beta}}(12)$ and the observable $(\bm{\sigma _3 \cdot c})$
commute because they act on different Hilbert spaces, which is not the case in the present one-particle problem.

In what follows we concentrate on the quantum mechanical process defined by the
expectation value
%%%%%%%%%%%%%%%%%%%%%%%%
\begin{equation}
\langle \hat{O} \rangle_{QM}
\equiv\langle \psi | \hat{P}^{\bm{a}}_{+}\hat{A}\hat{P}^{\bm{a}}_{+} |\psi \rangle,
\label{<PAP>}
\end{equation}
%%%%%%%%%%%%%%%%%%%%%%%%
where $\hat{O}=\hat{P}^{\bm{a}}_{+}\hat{A}\hat{P}^{\bm{a}}_{+} = \hat{A}\hat{P}^{\bm{a}}_{+}$ and the state of the system $| \psi \rangle$  will be taken to be that with spin up in the direction $\bm{n}$, i.e.,
%%%%%%%%%%%%%%%%%%%%%%%%
\begin{equation}
| \psi \rangle = |+ \rangle^{\bm{n}}\;.
\label{psi}
\end{equation}
%%%%%%%%%%%%%%%%%%%%%%%%

We shall compute $\langle \hat{O} \rangle_{QM}$ in QM following two different mathematical routes, and for each route we shall look for the corresponding HVM.
While the QM result cannot depend on the mathematical procedure,
we shall see that for the HVM the situation is more subtle.

We shall call {\em Route A} the evaluation of $\langle\hat{O}\rangle_{QM}$ of Eq. (\ref{<PAP>}) regarding it as the expectation value of the observable
$\hat{P}^{\bm{a}}_{+}\hat{A}\hat{P}^{\bm{a}}_{+}$
in the state $|+ \rangle^{\bm{n}}$, i.e.,
%%%%%%%%%%%%%%%%%%%%%%%%
\begin{equation}
\langle \hat{O} \rangle_{QM}
=\,^{\bm{n}}\langle + |\hat{O}|+ \rangle^{\bm{n}}\;.
\label{PAP Route A}
\end{equation}
%%%%%%%%%%%%%%%%%%%%%%%%

We shall denote by {\em Route B} the evaluation of $\langle\hat{O}\rangle_{QM}$ of Eq. (\ref{<PAP>}) regarding it as the expectation value of the observable $\hat{A}$ in the state obtained by letting the projector
$\hat{P}^{\bm{a}}_{+}$ act on the state
$|+ \rangle^{\bm{n}}$: i.e.,
the expectation value $\langle \hat{O} \rangle_{QM}$
of Eq. (\ref{<PAP>}) is computed
regarding $\hat{A}$ as the observable, the state being
%%%%%%%%%%%%%%%%%%%%%%%%
\begin{equation}
|\psi' \rangle
= \frac
{\hat{P}^{\bm{a}}_{+} |+ \rangle^{\bm{n}}}
{\sqrt{\;^{\bm{n}}\langle + |\hat{P}^{\bm{a}}_{+}|+ \rangle^{\bm{n}}}}
=|+ \rangle^{\bm{a}}.
\label{proj_state}
\end{equation}
%%%%%%%%%%%%%%%%%%%%%%%%
For the expectation value $\langle \hat{O} \rangle_{QM}$ we then have
%%%%%%%%%%%%%%%%%%%%%%%%
\begin{equation}
\langle \hat{O} \rangle_{QM}
= \;
^{\bm{a}}\langle +  | \hat{A} |+ \rangle^{\bm{a}}
\;\;
^{\bm{n}}\langle +  | \hat{P}^{\bm{a}}_{+}|+ \rangle^{\bm{n}} .
\label{PAP Route B}
\end{equation}
%%%%%%%%%%%%%%%%%%%%%%%%
A HVM for the observable and the state for each one of the two routes will be proposed below.

%%%%%%%%%%%%%%%%%%%%%%%%%%%%%%%%%%%%%%
\subsection{A HVM of successive measurements for one spin-1/2 particle}\label{hvm_spin}

In order to propose a HVM for each one of the routes indicated above,
we make the following assumptions:

a) In a HV space $\Lambda$, with $\bm{\lambda} \in \Lambda$, we define the HV ``value" of the various observables to be measured, or, in the nomenclature of Ref. \cite{fine}, the ``response function" (giving the $\bm{\lambda}$-determined responses to the measurement) of the observable, which takes on the eigenvalues of the QM observable in various domains of the HV space $\Lambda$.
The response functions will depend, in principle, on the ``settings" defining the observable and, at least initially in our discussion, on the state $|+ \rangle^{\bm{n}}$,
through a possible dependence of the various $\Lambda$ domains on the state.

b) We assume a normalized (non-negative) probability density
%%%%%%%%%%%%%%%%%%%%%%%%
\begin{equation}
\rho_{\bm{{\rm state}}}(\bm{\lambda})
\label{rho(lambda)}
\end{equation}
%%%%%%%%%%%%%%%%%%%%%%%
defined on $\Lambda$ and dependent, in general, on the state of the system.

It is possible to choose the response functions of a) and the density of b) so as to reproduce any QM expectation value in the present one-spin-1/2 problem
\cite{mermin}.
However, when comparing the integrands arising from HV modelling the two QM routes
indicated above we shall find a conflict of a similar nature as the one found in the previous section.

%%%%%%%%%%%%%%%%%%%%%%%%%%%%%%%%%%%%%%%%%%
\subsubsection{Route A}
\label{SpinRoute_A}
The state is given by Eq. (\ref{psi}) and is defined by the parameter $\bm{n}$. We shall thus write the HV probability density as
%%%%%%%%%%%%%%%%%%%%%%%%%%
\begin{equation}
\rho_{\bm{n}}(\bm{\lambda}).
\label{rho}
\end{equation}
%%%%%%%%%%%%%%%%%%%%%%%%

The observable is
$\hat{P}^{\bm{a}}_{+}\hat{A}\hat{P}^{\bm{a}}_{+}$.
We make the following assignment of response functions:
%%%%%%%%%%%%%%%%%%%%%%%%%%
\begin{subequations}
\begin{eqnarray}
\hat{A} &\Longrightarrow&
A_{\bm{n}}(\bm{\lambda}; \bm{a})
= \left\{%
\begin{array}{c}
a_{+}, \hspace{5mm}{\rm when}\hspace{5mm} \bm{\lambda}\in \Lambda_{+,\bm{a}}      \\
a_{-}, \hspace{5mm}{\rm when}\hspace{5mm} \bm{\lambda}\in \Lambda_{-,\bm{a}}
\end{array}
\right.
\label{A}
\\
\hat{P}^{\bm{a}}_{+} &\Longrightarrow&
\Pi_{\bm{n}} (\bm{\lambda};+,\bm{a})
= \left\{%
\begin{array}{c}
1, \hspace{5mm}{\rm when}\hspace{5mm} \bm{\lambda}\in \Lambda_{+,\bm{a}}      \\
0, \hspace{5mm}{\rm when}\hspace{5mm} \bm{\lambda}\in \Lambda_{-,\bm{a}}
\end{array}
\right.  \; ,
\label{spinPi}
\end{eqnarray}
\label{A,spinPi}
\end{subequations}
%%%%%%%%%%%%%%%%%%%%%%%%
which are allowed to depend on the ``settings" and, for the time being, on the state,
through a possible dependence of the domains $\Lambda_{\pm,\bm{a}}$ on $\bm{n}$.

Since the two QM operators in (\ref{A,spinPi}) {\em commute},
$[\hat{A}, \hat{P}^{\bm{a}}_{+}]=0$,
we assume, as in Ref. \cite{mermin}, that {\em the response function associated with their product is the product of the individual response functions} given in (\ref{A,spinPi}).

In terms of the HV assignments given above, we write the HVM image of the expectation value
(\ref{PAP Route A}) as
%%%%%%%%%%%%%%%%%%%%%%%%
\begin{subequations}
\begin{eqnarray}
\langle O \rangle_{HV}^{(A)}
&=&\int A_{\bm{n}}(\bm{\lambda};\bm{a}) \Pi_{\bm{n}}(\bm{\lambda};+,\bm{a})
\rho_{\bm{n}}(\bm{\lambda}) d\bm{\lambda} ,
\label{PiAPi Route A 1}
\end{eqnarray}
%%%%%%%%%%%%%%%%%%%%%%%%
which can also be written as
%%%%%%%%%%%%%%%%%%%%%%%%%
\begin{eqnarray}
\langle O \rangle_{HV}^{(A)}
&=&\int A_{\bm{n}}(\bm{\lambda};\bm{a})\rho_{\bm{n}}(\bm{\lambda}|+,\bm{a}) d\lambda
\int \Pi_{\bm{n}}(\bm{\lambda'};+,\bm{a})\rho_{\bm{n}}(\bm{\lambda'}) d\bm{\lambda'}.
\label{PiAPi Route A 2}
\end{eqnarray}
\label{PiAPi Route A}
\end{subequations}
%%%%%%%%%%%%%%%%%%%%%%%%%%%%
We have defined the {\em conditional probability density}
%%%%%%%%%%%%%%%%%%%%%%%%
\begin{equation}
\rho_{\bm{n}}(\bm{\lambda}|+,\bm{a})
=\frac
{\Pi_{\bm{n}}(\bm{\lambda};+,\bm{a}) \rho_{\bm{n}}(\bm{\lambda})}
{\int \Pi_{\bm{n}}(\bm{\lambda'};+,\bm{a})\rho_{\bm{n}}(\bm{\lambda'}) d\bm{\lambda'}} \; ,
\label{conditional}
\end{equation}
%%%%%%%%%%%%%%%%%%%%%%%%
which is the original $\rho_{\bm{n}}(\bm{\lambda})$ conditioned on $\bm{\lambda}\in \Lambda_{+,\bm{a}}$
[this is the domain where $A_{\bm{n}}(\bm{\lambda};\bm{a}) = a_{+}$;
see Eq. (\ref{A})].

%%%%%%%%%%%%%%%%%%%%%%%%%%%%%%
\subsubsection{Route B}
\label{SpinRoute_B}

The state is given by Eq. (\ref{proj_state}) and is defined by the parameter $\bm{a}$. We shall thus write the HV probability density as
%%%%%%%%%%%%%%%%%%%%%%%%
\begin{equation}
\rho_{\bm{a}}(\bm{\lambda}) \; .
\label{density-final-1spin}
\end{equation}
%%%%%%%%%%%%%%%%%%%%%%%%
To the observable $\hat{A}$ we assign the response function
%%%%%%%%%%%%%%%%%%%%%%%%
\begin{equation}
A_{\bm{a}}(\bm{\lambda};\bm{a}).
\label{response-fctn-final-1spin}
\end{equation}
%%%%%%%%%%%%%%%%%%%%%%%%

In terms of the HV assignments given above, we write the HVM image of the expectation value
(\ref{PAP Route B}) as
%%%%%%%%%%%%%%%%%%%%%%%%
\begin{equation}
\langle O \rangle_{HV}^{(B)}
=\int A_{\bm{a}}(\bm{\lambda};\bm{a})
\rho_{\bm{a}}(\bm{\lambda}) d\lambda
\int \Pi_{\bm{n}}(\bm{\lambda'};+,\bm{a})\rho_{\bm{n}}(\bm{\lambda'}) d\bm{\lambda'}.
\label{PiAPi Route B}
\end{equation}
%%%%%%%%%%%%%%%%%%%%%%%%

The HV expression $\langle O \rangle_{HV}^{(A)}$ of Eq. (\ref{PiAPi Route A 2})
was obtained from a {\em projection in the HV $\bm{\Lambda}$ space} unto the domain $\Lambda_{+,\bm{a}}$.
In contrast, the expression $\langle O \rangle_{HV}^{(B)}$ of
Eq. (\ref{PiAPi Route B})
was obtained by {\em first projecting in Hilbert space the QM state vector
unto $|+\rangle^{\bm{a}}$
and then finding the HV image}.
Both HVM expressions coincide with the QM result $\langle \hat{O} \rangle_{QM}$ and are therefore equal, so that we are fulfilling the requirement that a HVM should reproduce correctly the QM expectation values \cite{mermin}.
However, as we shall see,
we shall also find it interesting to study the detailed structure of the integrands in the two expressions (\ref{PiAPi Route A 2}) and (\ref{PiAPi Route B}).
We do this in what follows.

\underline{Response Functions}. First, we recall that $A_{{\rm state}}(\bm{\lambda};\bm{a})$ is an observable quantity, since it gives one of the individual results when measuring the observable $\bm{A}$:
at a given point in the HV $\bm{\Lambda}$ space, i.e., for a given $\bm{\lambda}$,
$A_{{\rm state}}(\bm{\lambda};\bm{a})$ takes on the value $1$ or the value $-1$,
and this is what shows up as an individual result of the measurement.
Now, we expect this value to be independent of the mathematical route that was followed to construct the HVM;
otherwise, the ``preassigned" value of the observable would be different in the two routes.
We recall that $A_{\bm{n}}(\bm{\lambda};\bm{a})$ of Eq. (\ref{A})
was allowed to depend on the state $|+\rangle^{\bm{n}}$, and
$A_{\bm{a}}(\bm{\lambda};\bm{a})$ of
Eq. (\ref{response-fctn-final-1spin})
was allowed to depend, through its lower index,
on the corresponding state $|+ \rangle^{\bm{a}}$.
But then in the second expression (i.e., $A_{\bm{a}}(\bm{\lambda};\bm{a})$) there is no longer a dependence on $\bm{n}$.
Also, suppose we restart our analysis from Eq. (\ref{<PAP>}) using the projector
$\hat{P}^{s_0\bm{a}}_{+} |\psi \rangle$, with $s_0=\pm 1$:
the upper index $\bm{a}$ in Eqs. (\ref{proj_state}) and (\ref{PAP Route B}) would then become $s_0\bm{a}$;
the lower index $\bm{a}$ in Eqs. (\ref{density-final-1spin}) and (\ref{response-fctn-final-1spin}) would also become $s_0\bm{a}$.
But then the response function of Eq. (\ref{response-fctn-final-1spin})
has an $s_0$ dependence, while that of
Eq. (\ref{A}) has no $s_0$ dependence.

For consistency, the two response functions
must be the same function of $\bm{\lambda}$ (and the setting $\bm{a}$) and this can only be achieved if they do not depend on the state, i.e.,
%%%%%%%%%%%%%%%%%%%%%%%%
\begin{equation}
A_{\bm{n}}(\bm{\lambda};\bm{a})
= A_{\bm{a}}(\bm{\lambda};\bm{a})
=A(\bm{\lambda};\bm{a}).
\label{A indep of state}
\end{equation}
%%%%%%%%%%%%%%%%%%%%%%%%
This result is an illustration of Eq. (\ref{C=C'}) that we found for the TP process.
As a result,
{\em when dealing with the expectation value of an observable which was preceded by a projective measurement (assuming commutation of the observable and the projector),
consistency of the notion of preassigned values requires the response functions not to depend on the state}.

Eq. (\ref{A indep of state}) is a simple illustration of the requirement
(\ref{C=C'}) found in the previous section.
We stress that no conflict arises when
just considering the resulting expectation value,
either of a single observable, or of the product of an observable and a projector, as in (\ref{<PAP>}).
In this latter case the conflict between state-dependent response functions and the notion of unique preassigned values arises when comparing the response functions that appear in the HVM of the QM expression that
(A) regards the projector as part of the observable, or
(B) as acting on the state to produce a new state.

Notice that Model 2 described in App. \ref{model 2} fulfills the property of having a state-independent response function.

\underline{HV densities}. The HV ``conditional" density
$\rho_{\bm{n}}(\bm{\lambda}|+,\bm{a})$ of Eq. (\ref{conditional})
was obtained from a projection in  HV space unto the domain
$\Lambda_{+,\bm{a}}$. In contrast, the ``final" density
$\rho_{\bm{a}}(\bm{\lambda})$ of Eq. (\ref{density-final-1spin})
was obtained by first projecting in Hilbert space the QM state
vector unto $|+\rangle^{\bm a}$ and then finding the HV image. In
what follows we prove (by contradiction) that  the conditional and
final densities are not the same. Where these densities the same,
we would have
%%%%%%%%%%%%%%%%%%%%%%%%%%%
\begin{equation}
\Pi(\bm{\lambda};+,\bm{a}) \rho_{\bm{n}}(\bm{\lambda})
=p \; \rho_{\bm{a}}(\bm{\lambda}) ,
\label{1_spin_cond=final}
\end{equation}
%%%%%%%%%%%%%%%%%%%%
where
%%%%%%%%%%%%%%%%%%%%%%%%%%%
\begin{eqnarray}
p=\int \Pi(\bm{\lambda};+,\bm{a})\rho_{\bm{n}}(\bm{\lambda})d\bm{\lambda}
= \; ^{\bm{n}}\langle + |P_{+}^{\bm{a}}|+ \rangle^{\bm{n}}
=\left| ^{\bm{a}}\langle + |+ \rangle^{\bm{n}}  \right|^2
\label{p 1}
\end{eqnarray}
%%%%%%%%%%%%%%%%%%%%%%%%%%
has been required all along to coincide with the QM probability --appearing in the last two expressions in (\ref{p 1})-- to find a positive projection in the direction $\bm{a}$ when the state is
$|+ \rangle^{\bm{n}}$.
Notice that if we integrate both sides of Eq. (\ref{1_spin_cond=final}) over $\bm{\lambda}$, we get the same result.

Suppose now that in (\ref{1_spin_cond=final}) we interchange
$\bm{a} \Leftrightarrow \bm{n}$.
Noticing, from the last expression in Eq. (\ref{p 1}), that $p$ does not change, we obtain
%%%%%%%%%%%%%%%%%%%%%%%%%%%%%
\begin{equation}
\Pi(\bm{\lambda};+,\bm{n})\rho_{\bm{a}}(\bm{\lambda})
=p \; \rho_{\bm{n}}(\bm{\lambda}).
\label{1_spin_cond=final 2}
\end{equation}
%%%%%%%%%%%%%%%%%%%%%%%
Notice that if we integrate both sides of Eq. (\ref{1_spin_cond=final 2}) over $\bm{\lambda}$, we get the same result.

For $p\neq 0$ we now solve Eq. (\ref{1_spin_cond=final}) for $\rho_{\bm{a}}(\lambda)$
and substitute it in (\ref{1_spin_cond=final 2}); we obtain
%%%%%%%%%%%%%%%%%%%%%%%%%%%%%
\begin{equation}
\Pi(\bm{\lambda};+,\bm{n})\Pi(\bm{\lambda};+,\bm{a})\rho_{\bm{n}}(\bm{\lambda})
=p^2 \; \rho_{\bm{n}}(\bm{\lambda}).
\label{1_spin_cond=final 3}
\end{equation}
%%%%%%%%%%%%%%%%%%%%%%%
But $\Pi(\bm{\lambda};+,\bm{n})\rho_{\bm{n}}(\bm{\lambda})
=\rho_{\bm{n}}(\bm{\lambda})$,
as it can be seen multiplying
both sides of (\ref{1_spin_cond=final 2}) by $\Pi(\bm{\lambda};+,\bm{n})$, using the fact that
$\Pi(\bm{\lambda};+,\bm{n})$ is idempotent and that $p\neq 0$.
So, (\ref{1_spin_cond=final 3}) becomes
%%%%%%%%%%%%%%%%%%%%%%%%%%%%%
\begin{equation}
\Pi(\bm{\lambda};+,\bm{a})\rho_{\bm{n}}(\bm{\lambda})
=p^2 \; \rho_{\bm{n}}(\bm{\lambda}).
\label{1_spin_cond=final 4}
\end{equation}
%%%%%%%%%%%%%%%%%%%%%%%
Integrating both sides over $\lambda$ we obtain
%%%%%%%%%%%%%%%%%%%%%%%%%%%%%
\begin{equation}
\int \Pi(\bm{\lambda};+,\bm{a})\rho_{\bm{n}}(\bm{\lambda}) d\bm{\lambda}
=p^2 \; \int \rho_{\bm{n}}(\bm{\lambda}) d\bm{\lambda}.
\label{1_spin_cond=final 5}
\end{equation}
%%%%%%%%%%%%%%%%%%%%%%%
>From (\ref{p 1}), the LHS is $p$, so that
%%%%%%%%%%%%%%%%%%%%%%%%%%%%%
\begin{equation}
p=p^2,
\label{p=p^2}
\end{equation}
%%%%%%%%%%%%%%%%%%%%%%%
which holds only for $p=0$ or $p=1$; for $p \neq 0,1$ we have a contradiction.

We thus conclude that for a HVM which, by definition, should reproduce any QM expectation value, the conditional and final densities as defined above
cannot be identical.

As a result, the physical significance of the two densities, within the
``reality" that a HV model tries to restore, is doubtful.
This situation
is of a similar nature as the
one indicated for the TP problem in the paragraphs
around Eq. (\ref{C=F}).
We conclude that projecting in HV space the HVM $\rho_{\bm{n}}({\bm \lambda})$ in order to obtain a conditional probability density, as in Route A, does not give the same result as projecting in Hilbert space first and then HV modeling, as in Route B.
I.e., the two operations --projection and HV modeling-- are not interchangeable.
Thus,
{\em the Hilbert space projection operation cannot be described in terms of and is in conflict with, the standard rules of statistics for the density in HV space}.

Model 1 described in App. \ref{model 1} does not fulfill the requirement of
Eq. (\ref{A indep of state}) discussed above.
An example that fulfills this requirement is model 2 of App. \ref{model 2}.
However, one can
verify the above theorem to exhibit explicitly
that the conditional and final densities are not the same.

We have thus illustrated, in this simple one spin-$1/2$ problem, the conflict that was found in the previous section for the more complex three-particle TP problem.

%%%%%%%%%%%%%%%%%%%%%%%%%%%%%%%%%%%%%%%%%%%%%%%%%%%%%%%
\section{Discussion and conclusions}
\label{discussion_conclusions}

In the present paper we have studied the possibility of underpinning the standard TP process for spin-$1/2$ particles with a LHVM.

An essential step in the TP process is a selective projective
measurement performed on particles 1-2 and the communication --by classical means-- of the result of the measurement to observer 3, so that the latter can select out of the full ensemble the subensemble described by one term of the the full state.
This projection is described by Eqs. (\ref{av_in_proj_state 1}) and (\ref{P_beta,betabar})
and explained in detail right above these equations.
When one tries to describe this projection and the resulting successive measurement process in terms of a HVM, it is found
[see the statements
around Eq. (\ref{C=F})]
that the HV density for the state projected in Hilbert space cannot be obtained manipulating the HV density for the original state following the rules of statistics in HV space:
i.e., {\em the Hilbert space projection operation cannot be described in terms of, and is in conflict with, the standard rules of statistics in HV space}.
Following two different mathematical routes (for the same experimental arrangement) we found two different HV densities; then the latter end up having a doubtful physical significance within the ``reality" that a HVM tries to restore.
This result we consider as a ``no-go theorem" for the HV description of the teleportation process.

Our considerations were based on HV models which guarantee the classical character of all the probabilities which can be deduced from them.
The assumptions made in our model were listed at the end of Sec. \ref{TPprocess}.

One might think of using a different HV model, which might be called the ``EV model", in which the eigenvalues of the observables are regarded as HV's, and the corresponding QM probabilities as HV weights.
In such a model, the conflict stated above would not arise:
no wonder, because one would just be doing QM in Hilbert space.
But, would such a model qualify as a HV model?
The idea behind introducing a HV model is precisely to have probabilities with classical properties at all stages, in order to restore local realism.
In the two-spin-$1/2$ problem studied originally by Bell, the probabilities involved in the EV model have classical properties when, for instance, the directions involved for the two measurements, $\bm{a}$ and $\bm{b}$, are parallel. The moment we extend the situation to non-parallel directions, we may encounter probabilities with non-classical features, in the sense that they violate the ``polytope inequalities" arising from classical notions of probability \cite{pitowsky}.
As we mentioned in the Introduction, in the three-spin-$1/2$ problem relevant for TP the correlations studied in Ref. \cite{zukowski} exhibit non-classical properties, and the same is thus expected for the related probabilities.

More generally, one may inquire about the possibility of generalizing the HV density
$\rho_{\psi}(\bm{\lambda})$ of (\ref{rho(lambda) TP}) to allow for a dependence of that density on the settings of the various observables to be mesasured \cite{brody}.
Take the two-spin-$1/2$ Bell problem again. One can show that
by doing so, even with local response functions one can
reproduce the QM probabilities, which, in turn, may violate the polytope inequalities;
such a setting dependence may thus produce probabilities of a non-classical nature.

In contrast, in the foregoing discussion we have been able to make specific statements for a class of HV models in which all of the probabilities which can be computed are guaranteed to be of a classical nature, like those arising from ``balls in an urn" \cite{pitowsky}.

The conclusion of our analysis that was stated above is crucially tied up with
assumption 3) listed at the end of Sec. \ref{TPprocess}: the splitting of
the HV $\bm{\lambda}$ in three sets,
$\{\bm{\lambda}_1, \bm{\lambda}_2, \bm{\lambda}_3 \}$, to be associated with each of the three particles, respectively.
This assumption was used starting from Eq. (\ref{rho(123)}) to represent states which, quantum-mechanically, are separable with respect to two subsystems;
we have not been able to deal with this feature without making such an assumption.
It is relevant to mention that this situation is reminiscent of the one that occurs when one uses Wigner's transforms to study TP for infinite-dimensional, non-denumerable, Hilbert spaces; the equivalent of each
$\bm{\lambda}_i$ is then the phase space of the $i$-th particle \cite{kalev_et_al}.
Needless to say, this point has to be investigated further.

\appendix

%%%%%%%%%%%%%%%%%%%%%%%%%%%%%%%%%%%%%%%%%
\section{HVMs for a single spin-1/2 particle}
\label{hvm_1_spin_1/2}

%%%%%%%%%%%%%%%%%%%%%%%%%%%%%%%%%%%%%%%%
\subsection{Model 1}
\label{model 1}

We first recall the HVM proposed by Bell \cite{bell 1} to describe the QM expectation value
%%%%%%%%%%%%%%%%%%%%%
\begin{equation}
^{\bm{n}}\left\langle + \right|
\bm{\sigma \cdot a}
 | + \rangle^{\bm{n}} = \bm{n \cdot a} \; ,
\label{<sa>n}
\end{equation}
%%%%%%%%%%%%%%%%%%%%%
associated with one spin-1/2 particle.
In that model, the HV $\bm{\lambda }$ has a uniform probability density over
the hemisphere $\bm{\lambda \cdot n}>0$, i.e.,
%%%%%%%%%%%%%%%%%%%%%
\begin{equation}
\rho ^{(Bell)}_{\bm{n}}(\bm{\lambda }) = \frac{1}{4\pi}
\left[1+{\rm sgn}(\bm{\lambda \cdot n})\right].
\label{bell_rho}
\end{equation}
%%%%%%%%%%%%%%%%%%%%%
Calling $\theta $ be the angle between $\bm{a}$ and $\bm{n}$, i.e.,
%%%%%%%%%%%%%%%%%%%%%
\begin{equation}
\cos\theta = \bm{a \cdot n}.
\label{theta}
\end{equation}
%%%%%%%%%%%%%%%%%%%%%
we define a unit vector $\tilde{\bm{a}}$,
obtained from $\bm{a}$ by a rotation of
$\bm{a}$ towards $\bm{n}$ until
%%%%%%%%%%%%%%%%%%%%%
\begin{eqnarray}
\tilde{\theta } = \frac{\pi}{2}(1-\cos \theta),
\label{theta_tilde}
\end{eqnarray}
%%%%%%%%%%%%%%%%%%%%%
$\tilde{\theta}$ being the angle between $\tilde{\bm{a}}$ and $\bm{n}$,
i.e.,
%%%%%%%%%%%%%%%%%%%%%
\begin{equation}
\cos\tilde{\theta }= \tilde{\bm{a}} \bm{\cdot n}.
\label{theta_tilde 1}
\end{equation}
%%%%%%%%%%%%%%%%%%%%%
The unit vector $\tilde{\bm{a}}(\bm{a},\bm{n})$ is thus a function of
both $\bm{a}$ {\em and} $\bm{n}$.
Now, let the result of an individual measurement of the component $\bm{\sigma \cdot a}$ be
defined by the ``response function" \cite{fine}
%%%%%%%%%%%%%%%%%%%%%
\begin{equation}
A_{\bm{n}}^{(Bell)}(\bm{\lambda },\bm{a})
={\rm sgn}(\bm{\lambda \cdot} \tilde{\bm{a}}) .
\label{bell_A}
\end{equation}
%%%%%%%%%%%%%%%%%%%%%
For its ensemble average we find
%%%%%%%%%%%%%%%%%%%%%
\begin{eqnarray}
\langle A^{(Bell)} \rangle
&=& \int A_{\bm{n}}^{(Bell)}(\bm{\lambda },\bm{a})
\rho ^{(Bell)}_{\bm{n}}(\bm{\lambda }) d\lambda
\nonumber \\
&=& 1- \frac{2\tilde{\theta}}{\pi} = \cos \theta,
\label{bell_<A> 1}
\end{eqnarray}
%%%%%%%%%%%%%%%%%%%%%
as required by Eq. (\ref{<sa>n})
[we have used Eq. (\ref{theta_tilde})].

%%%%%%%%%%%%%%%%%%%%%%%%%%%%%%%%%%%%%%%%%%%%%%%%%%
\subsection{Model 2}
\label{model 2}

The response function (\ref{bell_A}) for the observable
$\bm{\sigma \cdot a}$ has the peculiarity
of depending on the setting
$\bm{a}$ of the instrument {\em and} also on the state, defined by $\bm{n}$, on which the observable is measured.
The dependence of the observable on the state has been ruled out on physical grounds, when dealing with projectors, right after Eq. (\ref{C=C'}) and Eq. (\ref{A indep of state}).
This leads us to search for a model in which the response function is independent of the state.
In point of fact we can prove that the QM expectation value (\ref{<sa>n})
can be reproduced
using a density which depends {\em only} on the state $\bm{n}$ and a response function which depends {\em only} on the setting $\bm{a}$.
It is enough to choose
%%%%%%%%%%%%%%%%%%%%%
\begin{subequations}
\begin{eqnarray}
\rho _{\bm{n}}(\bm{\lambda })
&=& \frac{1+{\rm sgn}(\bm{\lambda \cdot n})}{2\pi}
(\bm{\lambda \cdot n})
\label{rho 1}
\\
A(\bm{\lambda },\bm{a})
&=& {\rm sgn}(\bm{\lambda \cdot}\bm{a}).
\label{A 1}
\end{eqnarray}
\label{rho,A 1}
\end{subequations}
%%%%%%%%%%%%%%%%%%%%%%%%%

\end{document}